# Nanoscale Probing of Localized Surface Phonon Polaritons in SiC Nanorods with Swift Electrons

*Yuehui Li[1,2], Ruishi Qi[1,2], Ruochen Shi[1,2], Ning Li[2], Yuanwei Sun[1,2], Bingyao Liu[2], Peng Gao[1,2,3]\**

[1]International Center for Quantum Materials, Peking University, Beijing, 100871, China.

[2]Electron Microscopy Laboratory, School of Physics, Peking University, Beijing, 100871, China.

[3]Collaborative Innovation Center of Quantum Matter, Beijing 100871, China.

\*Corresponding author. E-mail: p-gao@pku.edu.cn

# ABSTRACT

Surface phonon polaritons hold much potential for subwavelength control and manipulation of light at the infrared to terahertz wavelengths. Here, aided by monochromatic scanning transmission electron microscopy - electron energy loss spectroscopy technique, we study the excitation of optical phonon modes in SiC nanorods. Surface phonon polaritons are modulated by the geometry and size of SiC nanorods. In particular, we study the dispersion relation, spatial dependence and geometry and size effects of surface phonon polaritons. These experimental results are in agreement with dielectric response theory and numerical simulation. Providing critical information for manipulating light in polar dielectrics, these findings should be useful for design of novel nanoscale phonon-photonic devices.

# I. INTRODUCTION

Surface polaritons are electromagnetic surface modes formed by the strong coupling between light and electric or magnetic dipole-carrying excitations, which may result from collective conduction electron oscillations (surface plasmon polaritons, SPPs) or from lattice vibrations in polar crystals (surface phonon polaritons, SPhPs) [1]. Extensive works on SPP have proved it a promising candidate for nanoscale photonic circuits [2,3]. However, their inherent high optical loss severely constrains their applications. Many recent studies have been focused on SPhPs in polar dielectrics, which offer an alternative to achieve low-loss optical devices in the IR to terahertz (THz) spectral ranges [4]. The effective wavelength of SPhPs can be up to ten times shorter than the free-space wavelength, thus enabling nanoscale control of light far beyond the diffraction limit [5]. Compared with plasmonic materials, remarkably smaller imaginary part of the complex permittivity of phononic materials is advantageous for high Q resonances [6]. Previous works have revealed many of their desirable physical attributes, such as energy transfer [7], spatiotemporal coherent control of lattice vibration [8], and negative index [9]. However, due to difficulties in experimental detection, their geometry and size effects and subwavelength light confinement at the nanoscale remains elusive. Exploring the behavior of SPhPs in low dimensional systems [5,10,11] and nanomaterials [7,12,13], and further finding a way to effectively control and manipulate them, are therefore of vital significance.

Technically, several vibrational spectroscopy techniques including Raman scattering spectroscopy [14], infrared absorption spectroscopy [15], inelastic neutron scattering [16], inelastic X-ray scattering [17] and high-resolution electron energy loss spectroscopy [18] have a poor spatial resolution, precluding the investigation of size and geometry effects at the nanoscale. The spatial resolution of the tip enhanced Raman spectroscopy (TERS) and scanning near-field optical microscopy (SNOM) can recently be ameliorated to ~20 nm, while further improvement is limited by tip-sample contact [19]. To overcome these problems, atomic-wide electron beams have been used to probe localized SPhPs via electron energy loss spectroscopy (EELS) in a scanning transmission electron microscope (STEM) [20–22], which is equipped with a recently-designed monochromator and an ultra-bright cold field emission electron gun [23] and provides an electron probe with ~6 meV in energy resolution. This state-of-the-art facility allows us to spatially map localized SPhPs in a single nanostructure even down to the atom level and opens up unprecedented opportunities to delve into the unexplored field [23–31].

Here, we present EELS measurements of localized SPhPs in SiC nanorods, in which the localized SPhPs enable near-IR light confinement at the nanoscale. We reveal that the dispersion relation and exponential decay behavior in SiC nanorods are prominently dependent on the nanorod diameter. Furthermore, in nanorods with nonuniform diameter and shape, localized SPhP field can be concentrated on specific positions, indicating that localized SPhPs can be effectively controlled and manipulated by tailoring the geometry of the nanorod. These findings advance our understanding of localized SPhP behavior at the nanoscale and could help to better design polaritonic applications.

**II. METHODS**

The SiC nanorod investigated was a commercially available product (HEFEI KEJING MATERIALS TECHNOLOGY CO., LTD, HeFei, China). The SiC nanorods were ultrasoniced in alcohal for ~30 minutes and then were dispersed onto TEM sample grids (holy carbon file) in order to find isolated nanorods. All the EELS data were acquired on a Nion UltraSTEM$^{TM}$ 200 aberration-corrected electron microscope operating at 60 kV. The beam convergence semi-angle was 1.5 mrad and a slot aperture was used to capute special diffraction spots when acquiring the momentum-resolved EELS. In the Localized SPhPs EELS experiment, the beam convergence semi-angle was 15 mrad and the collection semi-angle was 24.9 mrad with a 1 mm spectrometer entrance aperture. The typical energy resolution (half width of the full zero loss peak, ZLP) was 8 meV for a high signal-to-noise ratio. The probe beam current was ~5-10 pA and the dispersion of per channel was 0.47 meV. The typical dwell time was 100-200 ms to achieve a satisfactory signal-to-noise ratio. Gatan Digital Microscopy software and MATLAB were used to process the data. Atomic-resolved high angle annular dark field (HAADF) images were acquired using FEI Titan Themis operating at 300 kV.

Due to the instability of the electron beam, the shape of ZLP may change slightly over time. To correct the energy shift of ZLP, we aligned the center of ZLP to zero. To eliminate the influence of beam fluctuation, the aligned spectra were normalized using unsaturated counts of the ZLP. In the loss region, the ZLP forms a background, which was fitted using the power low $I(E) = I_0 E^{-r}$ ($I_0$ and $r$ are adjusted parameters) and then subtracted [32]. Numerical calculations were based on boundary element method (BEM) via the open MATLAB MNPBEM Toolbox [33]. The simulations help us to caputre insight into localized SPhPs by accounting for the surface and shape effects. The toolbox incorporates dipole scattering and surface contributions that are in good agreement with experiments.

The dielectric function of SiC is calculated by the Lorentz oscillator model

$$\varepsilon(\omega) = \varepsilon_\infty \left[ 1 + \frac{\omega_L^2 - \omega_T^2}{\omega_T^2 - \omega^2 - i\Gamma\omega} \right] \quad (1)$$

where $\varepsilon_\infty = 6.7$ is the permittivity for high frequencies, $\omega_T = 793 cm^{-1}$ and $\omega_L = 969 cm^{-1}$ are the frequency of the transverse optical phonon and longitudinal optical phonon, respectively, and $\Gamma = 4.76 cm^{-1}$ is the damping constant [34].

### III. RESULTS AND DISCUSSION

The high angle annular dark field (HAADF) images in Fig. 1(a) show the morphology and atomic structure of a typical cubic SiC nanorod viewed from [110]. Figure 1(b) presents the phonon dispersion measured along the direction ΓRΓRΓ. Corresponding calculated dispersion line is also shown in the high-order Brillouin zone. The dispersion relation is obtained from the Materials Project (www.materialsproject.org) [35] and the phonon band paths are determined by using Seek-path developed by Giovanni Pizzi [36]. The calculated dispersion relation is in good agreement with the experiment in the high-order Brillouin zone, except that the TA was not obtained. The phonon dispersion was not so clear in the first and second Brillouin zone due to the broadening of the intense central bragg diffraction spot.

It is well known that the Reststrahlen band lies between the transverse optical (TO) phonon and longitudinal optical (LO) phonon frequencies, which are identified at the crossover of the real part (Re[$\varepsilon$] = 0). The termination surface breaks the translational symmetry and changes the dielectric environment of the bulk crystal, which leads to the generation of a third resonance [21], such as Fuchs-Kliewer mode (surface optical mode of infinite flat plane, Re[$\varepsilon$] = −1), Fröhlich mode (surface polariton of sphere, Re[$\varepsilon$] = −2) and SPhPs. The vibrational frequencies of SPhPs are limited in the Reststrahlen band, the permittivity in which is negative. From the point view of dispertion, the phonon polariton is formed near Γ [Figure 1(c)] due to interactions between the electromagnetic wave induced by fast electron beam and the TO phonon. Figure 1(d) displays a typical bulk phonon polariton dispersion (black line) and surface phonon polariton dispersion (cyan line) [37]. The latter was obtained by measuring the vibrational EELS along the SiC nanorod in this letter. The SPhP dispersion line lies between the transponse optical brach ($\omega_{TO}$) and surface phonon branch ($\omega_{SO}$).

The SPhPs EELS spectra were recorded in aloof geometry, which means that the electron beam passed near the nanorod but without intersecting it. In this case, the signal of bulk phonon, which is not of our interest, is excluded. Figure 1(e) displays typical EELS spectrum collected from the red region of a SiC nanorod [Figure 1(a)] with 240 nm in diameter and 2488 nm in length. The calculated spectrum convoluted with a Gaussian function with a width of 10 meV, which accounts for the instrument response, is in good agreement with the experimental EELS. To preclude the influence of carbon grid, the EELS signal of carbon grid that was recorded with the same experimental conditions is shown in Fig. S1 as a reference, in which no distinguiable signal was obtained in the energy region 80-180 meV. The simulated spatial distribution of inelastic electron scattering probability associated with different modes are presented in Fig. 1(f), showing the interference of localized SPhPs and consequent standing wave patterns. Localized SPhP modes excited by the long-range Coulomb field induced by the swift electrons [38–41] in a nanorod of length $L$ are similar to localized SPhP modes of an infinitely-long rod, except that the available wave vectors are confined to a multiple of $1/2L$ [12]. With $m$ denoting the number of nodes along the rod, the signal intensity concentrated at the ends of the nanorod mainly derives from the $m=1$ mode. With increasing order $m$, the wavelength reduces leading to higher resonant energy. What we measured in Fig. 1(g), the two dimensional plot of EELS as a function of beam position along the long axis of the nanorod, is the connection of corresponding anti-nodes (shown by the red dashed line) due to the limited experimental energy resolution. As a reference, the two dimensional plot of calculated EELS spectrum and convoluted spectrum are shown in Fig. S2.

Given a semi-infinite nanorod, the interference wave vector is continuous ($1/2L \to 0$), so we can get the dispersion of SPhPs in cylinders by converting the EELS data from coordinate space to reciprocal space. For a given position, the wave number $k = \pi/d_{\text{EELS}}$ where $d_{\text{EELS}}$ is the distance between the given position and the end of the nanorod, as illustrated in Fig. 2(b). Following this relation, we obtain the dispersion relation of SPhPs [Fig. 2(d)] of the semi-infinite nanorod in Fig. 2(a), derived from Fig. 2(c), which is in excellent agreement with the calculated dispersion of an infinite SiC cylinder with the same diameter (shown by pink dotted line, derived from Ref [42]). The dispersion relation for surface phonon polariton of cylinder boundary by J. C. Ashley and L. C. Emerson [42], written in the form:

$$v^{'}\varepsilon\alpha - v\varepsilon^{'}\beta = 0 \qquad (2)$$

where $\varepsilon$ ($\varepsilon^{'}$) is the dielectric function inside (outside) the cylinder, and

$$v \equiv \left(k^2 - \varepsilon \omega^2 / c^2\right)^{\frac{1}{2}} \tag{3}$$

$$v' \equiv \left(k^2 - \varepsilon' \omega^2 / c^2\right)^{\frac{1}{2}} \tag{4}$$

$$\alpha \equiv \frac{d}{d(va)} \ln I_0(va) \tag{5}$$

$$\beta \equiv \frac{d}{d(v'a)} \ln K_0(v'a) \tag{6}$$

where $c$ is the speed of light in vacuum, $a$ is the radius of cylinder, $k$ is the wavevector, $\omega$ is the vibrational frequency, the function $I_0$ and $K_0$ are the modified Bessel functions of the first and second kind of order zero. The dispersion is dependent upon both the dielectric function and the radius. For the limiting case of the plane interface, $a \to \infty$, $\alpha \to 1$, $\beta \to -1$, the surface phonon dispersion relation reduces to

$$v'\varepsilon + v\varepsilon' = 0 \tag{7}$$

Figure 2(e) shows the spectra acquired in nanorods of different radius $r$ at fixed wave number 0.005 nm$^{-1}$. Figure 2(f) shows the dispersion relations of SPhPs in SiC cylinders of different radii. The solid lines are theoretically calculated [42] and the solid scatters are our experimental data. As we expected, for larger values of radius $r$, the dispersion relation shows an incipient convergence to the curve for an infinite, flat surface (solid black line). At larger radius $r$ and larger wave number $k$, the curve approaches the energy of surface phonon $\omega_S$, determined by the equation $\varepsilon(\omega_S) + 1 = 0$ [38], which is the so-called FK mode [43]. The curve of dispersion relation all lies to the right of the light line (dash dot line), which means the localized SPhPs are nonradiative [44]. At low $k$, the group velocity of the localized SPhPs, determined by the slope, approaches the speed of light, as shown in inset of Fig. 2(f).

As evanescent waves highly localized at the surface, localized SPhP field decays exponentially along the direction perpendicular to the surface [45]. As shown in Fig. 2(g), we obtain the evanescent field data of different modes by selecting the EELS data projected along the long axis at corresponding simulated resonant energy. Then, the data was fitted with an exponential function of the form $I = I_0 \exp(-d/\lambda_d)$, where $\lambda_d$ is the decay length that describes the distance from the edge to the position where the localized SPhP field decreases by a factor of $1/e$, and $d$ is the distance from the nanorod surface. The inset in Fig. 2(g) displays the "decay length" of different order, where the red (black) scatter is based on the experiment (BEM simulation) data. As mode order and energy increase, the wavevector increases, leading to lower "decay length". The experiment data is in good agreement with the simulation data, except that the decay length of the low order resonances of the simulation is a bit larger than that of the experiment. This discrepancy may be resulted from the contribution of the field of higher modes to the measurement of the lower modes due to the limited energy resolution.

Figure 3(a) shows a sample whose geometry is approximately a conjunction of a cylinder and a cone. Figure 3(b) displays the corresponding two-dimensional plots of normalized intensity. The EEL probabilities of the sample are simulated in Fig. 3(c), which are convolved with a Gaussian function to account for non-ideal energy resolution. The simulation is in good agreement with the experiment. The symmetric distribution of the resonance mode of nanorod [Fig. 2(a)] is broken due to the asymmetric geometry with one sharp and the other flat ends of the nanorod. Figure 3(e) displays the two-dimensional plot of normalized intensity, corresponding to the sample shown in Fig. 3(d), like two back-to-back sandglasses. The interference due to the convex-edge reflection gives rise to higher-order resonance modes compared to the center axis, which is similar to the sample above. The smaller radius constrains the vibrational energy, as shown in Fig. 2(f), which leads to lower energy than the nanorod adjacent. Figure 3(f) shows the simulated EEL probabilities, which is in good agreement with Fig. 3(e). The near-field distribution at three energies 99, 106, and 117 meV were presented in Figs. 3(g), 3(h) and 3(i). The near field distribution at different energies is rather different. At 99 meV, the field amplitude at the nanorod is uniform. There is no constructuive or destructive interference as it approached the long wavelength limit. At 106 meV, however, there is constructive interference at the slim part, resulting in an increase in the near-field intensity. At 117 meV, the near-field intensity is high at the convex-edge. We attribute it to the high $k$ value owing to the interference between electron source and the convex-edge.

Figure 4(a) presents the simulated EEL probability surrounding the SiC nanorod in Fig. 3(d) at resonance energy. Figures 4(b) and 4(c) show the simulated electric field distribution in the vicinity of SiC excited by a planewave excitation with polarization along the axis and perpendicular to the axis, respectively, where the incident wavelengths were determined by the resonance energy in Fig. 4(a). The distribution of corresponding surface charge polarity is also set up across the nanorod. From these maps and comparing with those shown in Fig. 3(g), as expected, the EEL probability, surface charge density and electric field all show a strong confinement at the slim part. This is similar to "hot spot" in SPP, which appears at the point where the translation symmetry is broken [46]. This phenomenon may be utilized to concentrate electric field at the subwavelength scale.

The response differs between the electron and optical excitations. For the electron beam, both odd modes and even modes can be excited. However, for the plane wave whose electrical polarization is aligned with the nanorod axis, only odd modes (mode 1, 3, 5, 7) can be excited due to the selection rules, while even modes (mode 2, 4, 6, 8) cannot [12] without regard to retardation effects. The longitudinal modes cannot be excited when using polarized plane wave with polarization perpendicular to the nanorod axis. As shown in Fig. 4(c), no obvious field distribution similar to dipole modes was obtained. The distribution of electric fields with different orientation was shown in Fig. S3, which further demonstrates that only odd modes can be excited. Contrary to optical methods, EELS is sensitive to all modes. Therefore, besides higher spatial resolution, the atomic-size electron probe also overcomes the limits on coupling, polarization suffered by light-based techniques, being a powerful method to study phonon polaritons in nanostructures. So far there are still difficulties in distinguishing discrete modes due to the insufficient energy resolution, which should be solved with further improvement of STEM-EELS resolution in future.

## IV. CONCLUSION

In summary, we visualize the surface phonon polariton in SiC nanorods by STEM-EELS. We demonstrate that the electron beam excites localized SPhPs near the sample edge, and further show that by converting the coordinate space to the reciprocal space, the dispersion relations of SPhPs can be obtained in STEM-EELS systems. We show that the exponential decay of the field and its decay length is obtained for different modes. More importantly, by mapping the surface phonon polaritons using electron beam, our work demonstrates that STEM-EELS provides an approach to study the localization and dispersion relation of surfrace phonon polaritons in complex dielectric nanostructures.

**AUTHOR INFORMATION**

**Corresponding Author**

*E-mail: p-gao@pku.edu.cn

**Notes**

The author declare no competing financial interest.

**ACKNOWLEDGEMENTS**

We gratefully acknowledge the support from the National Program for Thousand Young Talents of China and "2011 Program" Peking-Tsinghua-IOP Collaborative Innovation Center of Quantum Matter. The authors acknowledge Electron Microscopy Laboratory of Peking University for the use of Cs corrected electron microscope. **Funding:** The work was supported by the National Natural Science Foundation of China (Grant Nos. 11974023, 51672007), the National Key R&D Program of China (2016YFA0300804), the National Equipment Program of China (ZDYZ2015-1), and Key Area R&D Program of Guangdong Provience (2018B010109009, 2018B010109009).

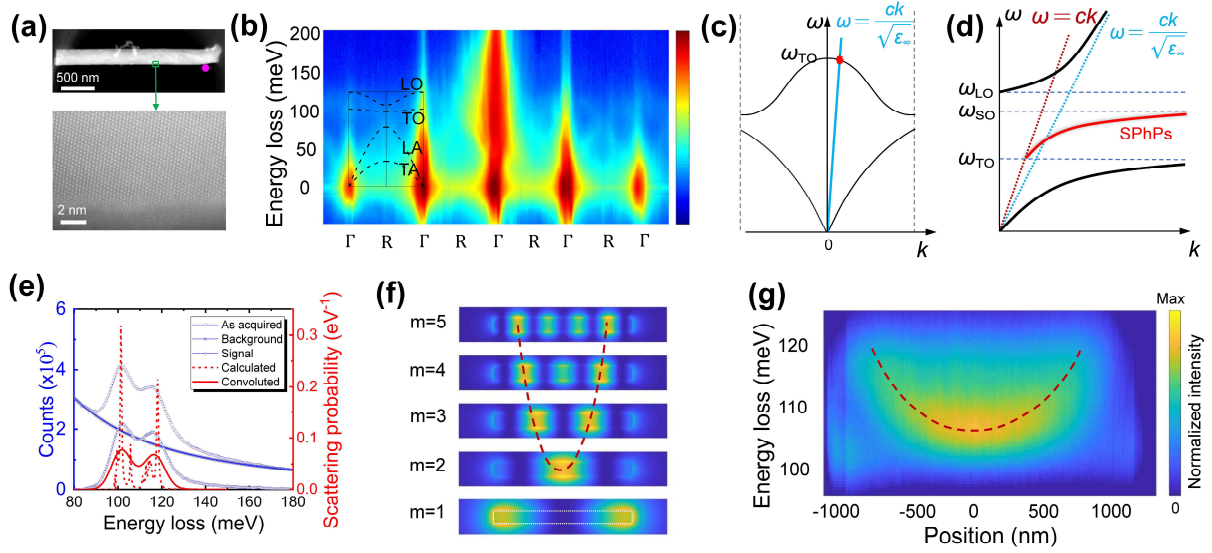

**Fig. 1. STEM-EELS measurement of phonon dispertion and and localized SPhPs in a SiC nanorod.** (**a**) HAADF image showing the morphology and atomic structure of a cubic SiC nanorod viewed from [110]. (**b**) Momentum-resolved EELS along the direction ΓRΓRΓ. The dispersion relation is obtained from the Materials Project (www.materialsproject.org) [35]. (**c**) The schematic of the coupling between the electromagnetic wave and transverse optical phonon. (**d**) The schematic of the dispersion of phonon polaritons [37]. (**e**) The typical phonon polariton signal obtained by experiment and simulation. The red dash line was the simulated EELS probabilities, containing several resonance modes clearly. The red solid line was the calculated spectrum convoluted with a Gaussian function with a width of 10 meV, which accounts for the response function. (**f**) Simulated spatial distribution associated with different modes. (**g**) Two dimensional plots of EELS.

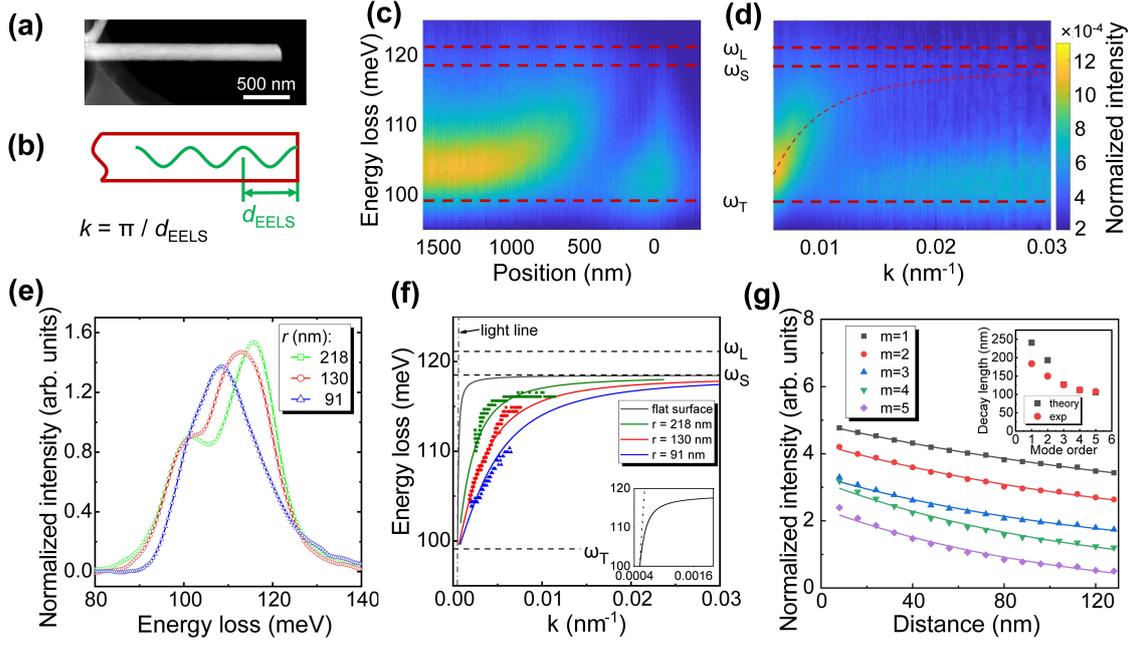

**FIG. 2. Spatial distribution and dispersion relation for different modes.** (**a**) HAADF image for a semi-infinite SiC nanowire. (**b**) Schematic diagram of interference. (**c**) Two dimensional plot of EELS. (**d**) Dispersion relation obtained from experiment data. The dotted line is calculated by dielectric theory. (**e**) EELS signal of nanorods with different radii at specific momentum transfer $0.005$ $\text{nm}^{-1}$. (**f**) Dispersion relation of SPhPs in nanorods with different radii. The solid line and scatters are obtained from theory and experiment, respectively. (**g**) Signal intensity of several localized SPhP modes as a function of distance to the surface. Scatters and solid curves are experimental data and exponential fitting respectively. Inset shows their theoretically calculated and experimentally fitted decay length.

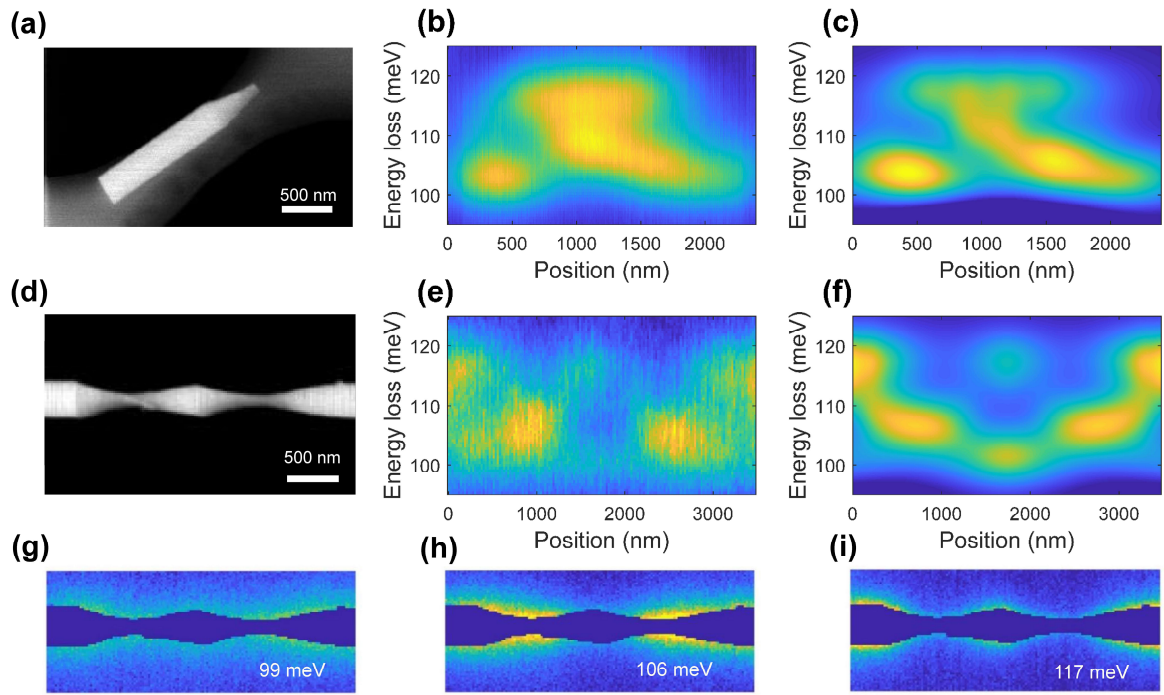

**FIG. 3. Geometry effects.** (**a**) HAADF image, (**b**) experimental EELS signal, and (**c**) simulated EELS of a nanorod with tip-end. (**d**) HAADF image, (**e**) experimental EELS signal, and (**f**) simulated EELS of a nanorod with back-to-back sandglasses. (**g, h, i**) Spatial distribution of intensity associated with different energies.

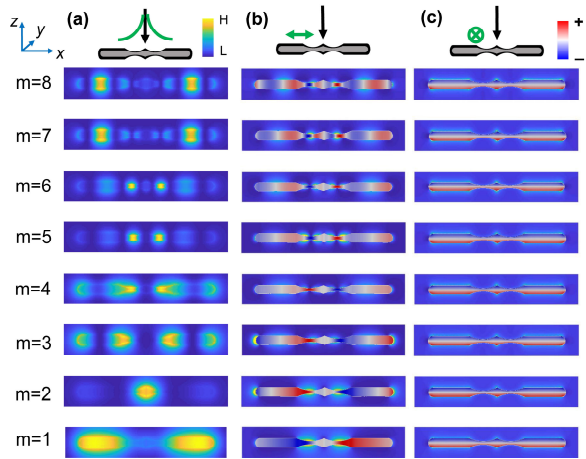

**FIG. 4. The comparison of simulated electric field distribution for electron and polarized light excitation schemes.** (**a**) Simulated EELS probability maps. (**b**, **c**) Simulated surface charge and electric field distribution $|E_x|^2$ under electric polarization along the axis (b) and perpendicular to the axis (c).

# Supplemental Material

# Nanoscale Probing of Localized Surface Phonon Polaritons in SiC Nanorods with Swift Electrons

*Yuehui Li[1,2], Ruishi Qi[1,2], Ruochen Shi[1,2], Ning Li[2], Yuanwei Sun[1,2], Bingyao Liu[2], Peng Gao[1,2,3]\**

[1]International Center for Quantum Materials, Peking University, Beijing, 100871, China

[2]Electron Microscopy Laboratory, School of Physics, Peking University, Beijing, 100871, China

[3]Collaborative Innovation Center of Quantum Matter, Beijing 100871, China

\*Corresponding author. E-mail: p-gao@pku.edu.cn

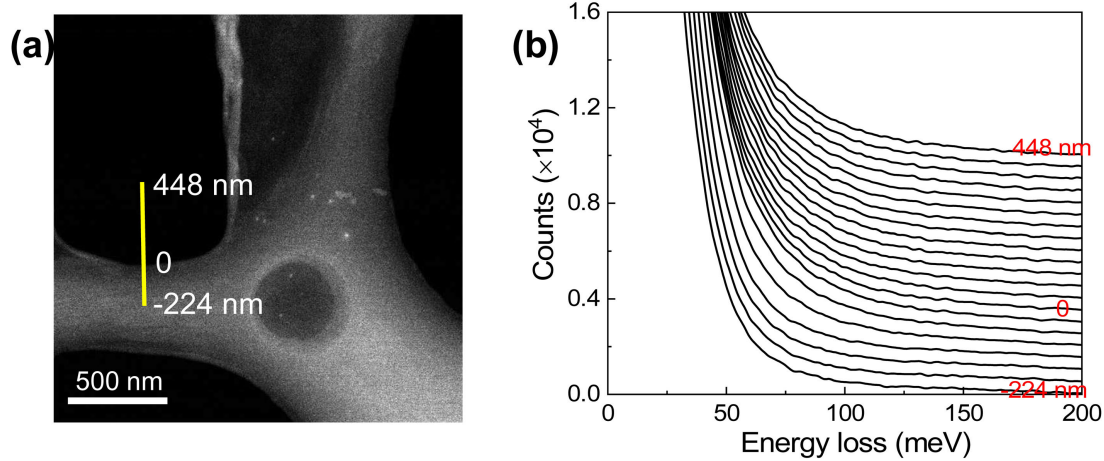

**FIG. S1. EELS of carbon grid.** (**a**) HAADF image of carbon grid. (**b**) EELS of carbon grid. No vibrational signal was obtained in the energy region 80-180 meV.

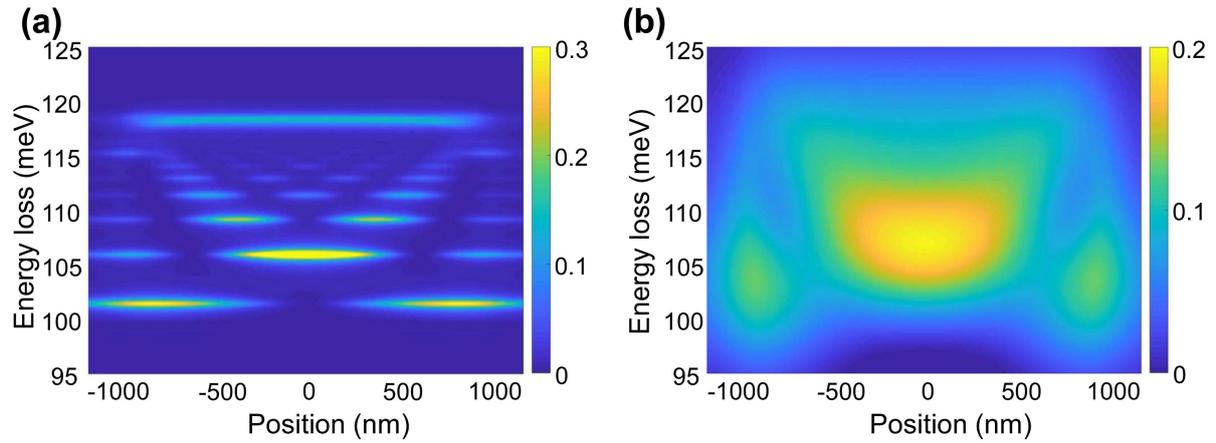

**FIG. S2. Simulated EELS probabilities.** (**a**) The simulated EELS probabilities corresponding to a SiC nanorod with radius 100 nm, length 2000 nm and impact parameter (distance from the cylinder surface) 100 nm. The interference pattern of different mode was shown clearly. (**a**) The EELS signal corresponding to the convolution of the (a) with a Gaussian function with a width of 10 meV.

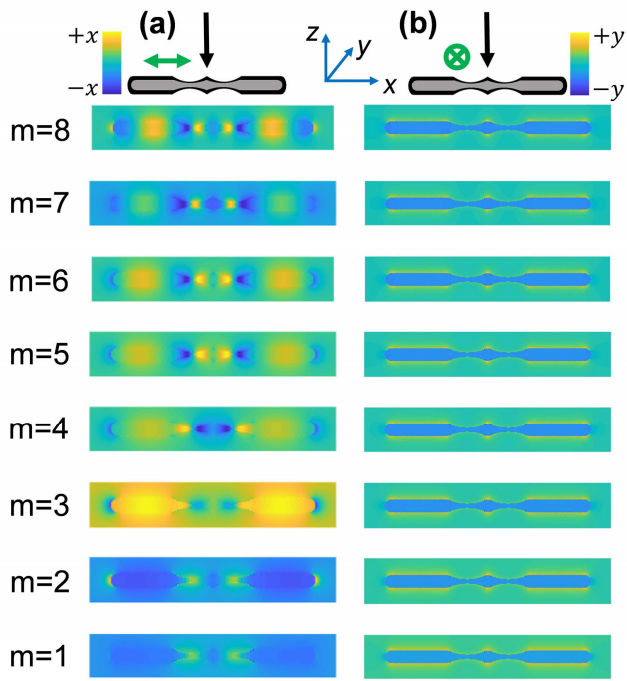

**FIG. S3. Spatial distribution of electric field.** (a) The $E_x$ component of electric field distribution under plane wave with polarization along the axis. (b) The $E_y$ component of electric field distribution under plane wave with polarization perpendicular to the axis.